\documentclass{article}
\usepackage{spconf,amsmath,graphicx, multirow, url}

\usepackage{mathtools}
\usepackage{blindtext}
\usepackage{caption}

\title{Detection of COVID-19 through the analysis of vocal fold oscillations}
\name{Mahmoud Al Ismail \qquad Soham Deshmukh \qquad Rita Singh}
\address{Carnegie Mellon University, Pittsburgh, PA, USA \\
         \texttt{\{mahmoudi,sdeshmuk,rsingh\}@andrew.cmu.edu}}
\begin{document}
%\ninept
%
\maketitle
\begin{abstract}
%Abstract should not have citations The vibration of the vocal folds is the primary source of vocalisation \cite{titze2008nonlinear}
Phonation, or the vibration of the vocal folds, is the primary source of vocalization in the production of voiced sounds by humans. 
%It is a self-sustained bio-mechanical process that is governed by the intricate balance of aerodynamic forces across the glottis, and is highly sensitive to changes in the speaker's respiratory parameters. 
It is a complex bio-mechanical process that is highly sensitive to changes in the speaker's respiratory parameters.
Since most symptomatic cases of COVID-19 present with moderate to severe impairment of respiratory functions, %including those of the lungs and musculature of the respiratory tract, 
we hypothesize that signatures of COVID-19 may be observable by examining the vibrations of the vocal folds. Our goal is to validate this hypothesis, and to quantitatively characterize the changes observed to enable the detection of COVID-19 from voice. For this, we use a dynamical system model for the oscillation of the vocal folds, and solve it using our recently developed ADLES algorithm to yield vocal fold oscillation patterns directly from recorded speech. Experimental results on a clinically curated dataset of COVID-19 positive and negative subjects reveal characteristic patterns of vocal fold oscillations that are correlated with COVID-19. We show that these are prominent and discriminative enough that even simple classifiers such as logistic regression yields high detection accuracies using just the recordings of isolated extended vowels.  

% From what is understood of the pathogenesis of Covid-19 until now, one of its wide-ranging symptoms is the  impairment of respiratory function in affected persons [CITE HERE]. Since voice production is driven by the upper respiratory tract, it stands to reason that Covid-19 related changes to its functions must also influence voice. This paper is based on the hypothesis that in cases where Covid-19 affects the upper respiratory tract, the motion of the vocal folds during speech production may also be compromised. If so, this is highly likely to influence the pattern of displacements and velocities of the individual vocal folds as they oscillate during phonation, and in the manner in which their vibrational eigen-frequencies lock during this process. This paper is exploratory in nature, and uses a recently proposed novel computational mechanism \cite{wenbo} to directly infer the motion of the vocal folds during phonation from recorded speech signals. We compare the inferred vocal fold parameters and motions across groups of speakers who are and are not symptomatic with Covid-19, using carefully curated data collected under expert supervision. Results show distinctive biomarkers of Covid-19 on vocal fold movements, which can be clearly differentiated from those of symptom-free persons. To quantify the visual analysis, we use interpretable simple classifier which can subsequently detect the presence of Covid-19 from parameters characterising the vocal fold oscillations of patients.

\end{abstract}

\vspace{-0.05in}
\section{Introduction}
\label{sec:intro}
The vibration of the vocal folds is the primary source of voicing (or \textit{phonation}) in humans \cite{titze2008nonlinear}. The membranes that comprise the vocal folds are partially tethered by the muscles, cartilage and ligaments surrounding them, allowing them to open and close the glottal area, and to vibrate in response to the passage of air through the glottis. As a result of their structure and physical placement in the larynx, they have characteristic eigen-modes of vibration, or eigen-frequencies at which they can independently vibrate. These are a function of the bio-physical properties of the vocal folds, such as their length, thickness, elasticity etc. During phonation, the vibrations of the two vocal fold membranes \textit{synchronize} or \textit{lock} at one of their many eigen-frequencies. 
%This vibration is mediated by the muscles of the larynx, but in itself is self-sustained. Both, the self-sustained 
Both, the oscillations of the vocal folds during phonation,  and this \textit{entrainment} (or synchrony during vibration), result from an intricate balance of aerodynamic forces across the glottis. These forces are directly dependent on the respiratory functions of the speaker, among other factors \cite{singh2019production}, and are highly sensitive to changes in them. The oscillation patterns of the vocal folds, the symmetry of their motion as the glottis opens and closes, the frequencies at which they synchronize (or the extent of their synchrony), can all be very easily compromised by fine fluctuations in the airflow dynamics of the upper respiratory tract, or even by slight impairments of any of the laryngeal muscles. Disturbances in any of these factors can cause the vocal folds to vibrate in an asymmetrical and asynchronized fashion, and to fail to lock due to unstable eigen-modes.

Clinical observations of symptomatic patients of COVID-19 have so far revealed that this virus moderately or often seriously impairs the functions of the lower and mid respiratory tract, including that of the lungs, airways and musculature of the respiratory tract. Patients who are symptomatic and have tested positive for COVID-19 as the underlying cause have not only reported changes in their voice, but also a general inability to \textit{produce} voice normally.  This leads us to hypothesize that the vocal folds of these persons are likely to exhibit anomalies in their oscillation patterns during phonation, and that these can be used to detect COVID-19 from voice. The goal of this paper is to validate this hypothesis.

\vspace{-0.1in}
\subsection{Related Work}
\label{sec:format}

%The study reported in this paper began concurrently with data collection shortly after the onset of the COVID-19 pandemic. At the time, not much was understood about the pathogenesis of COVID-19, let alone its influence on voice. 
As of now, literature on detecting COVID-19 from voice, coughs and other respiratory sounds is recent and sparse \cite{tcs_2020overview}. One study \cite{mit} has attempted to detect COVID-19 by analyzing the speech envelope, pitch, cepstral peak prominence and the formant center-frequencies. This study observes high-rank eigen-values tending toward relatively lower energy in  post-COVID-19 cases, but does not  provide strict interpretations. Researchers have also used crowd-sourced data \cite{brown2020exploring, Imran_2020} with data-driven end-to-end deep learning methods for this purpose. However, the data remain scarce, and deep learning models are prone to over-fitting -- there is no guarantee that the network will specifically learn only COVID-19 related characteristics, and not speaker-specific characteristics.  

A controlled medical study that is of special relevance to our work is reported by Huang \textit{et. al.} \cite{Huang2020}, which uses stethoscope data from lung auscultation to analyze the breathing patterns of COVID-19 patients. In this study, recorded audio signals were analyzed by six independent physicians. All COVID-19 patients were observed to have abnormal breath sounds like crackles, asymmetrical vocal resonances and indistinguishable murmurs. These results were reported to be consistent with CT scans of the $9^{th}$ intercostal cross-section of the corresponding patient. The study found concrete evidence of the association of abnormal breath sounds, and asymmetries in vocal resonances with COVID-19 infection. This study suggests that COVID-19 affects the source signal that excites the vocal tract, which implicates abnormalities in vocal fold oscillations. While it supports our hypothesis that observing vocal fold oscillations may yield information relevant to detection of COVID-19, it is infeasible to make such direct observations patient symptoms (using a stethoscope, or using high-speech videography of vocal fold motion) at scale for widespread diagnostic purposes.
% and close contact with the patient.

%However, obtaining data from lung auscultation is not a scalable and widely accessible testing method for COVID-19. It requires expert clinicians, appointments, personal interactions, and also requires the individuals to have ready accessibility to medical services.  

In our work, we use the much more scalable and accessible approach of computationally deducing the oscillations of the vocal folds directly from recorded speech signals. The algorithmic details of this approach are given in Sec. \ref{sec:2}. Experiments on  clinically curated data reveal the presence of clear bio-markers of COVID-19 in the vocal fold oscillation patterns, in the estimated glottal flow, and in the residuals obtained. In Sec. \ref{sec:3} we discuss these, and analyze their usefulness in detecting COVID-19 using multiple classifiers.

\section{Estimating vocal fold displacements}\label{sec:2}

\vspace{-0.05in}
\subsection{The vocal fold oscillation model}
\label{sec:vfo}

Of the several mathematical models of phonation proposed in the past decades \cite{1fold_asym, lucero2013modeling, ishizaka1972synthesis, yang2011computation, alipour2000finite, titze1988physics}, the 1-mass asymmetric body-cover model \cite{1fold_asym} is of particular interest to us due to its ability to capture asymmetry in the oscillation of left and right vocal folds. We briefly describe this model below.

\begin{figure}[!htb]
\begin{minipage}[b]{1.0\linewidth}
  \centering
  \centerline{\includegraphics[width=5cm]{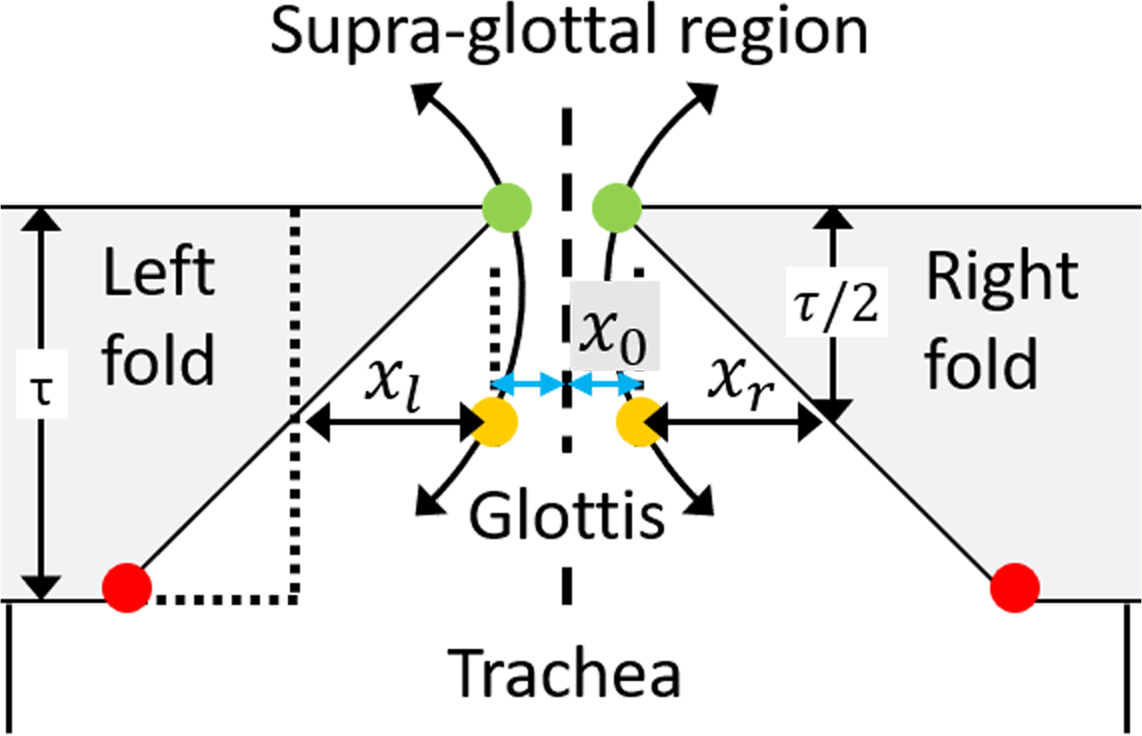}}
\end{minipage}
\caption{Schematic diagram depicting a cross sectional (frontal) view of the vocal folds. The folds have both horizontal and vertical (curved arrows) modes of oscillation. %The colored dots show symmetrically located points on the folds.
}
\vspace{-0.1in}
\label{fig:schematic}
\end{figure}

Fig. \ref{fig:schematic} shows a schematic diagram of the vocal folds. As they vibrate, the horizontal displacements of the left and right vocal folds ($x_l$ and $x_r$) are measured with reference to the center of the glottis (central dashed line). %The subscripts $l$ and $r$ indicate the left and right vocal folds respectively. 
$x_0$ represents displacements at rest.   The model measures the displacements at the location (yellow dots) where the folds are half their maximum thickness ($\tau$). The length of the vocal folds $d$ is normal to the plane of the figure and not shown.

The asymmetric 1-mass body-cover model is described by the set of coupled non-linear differential equations: 
\begin{align}
    &\ddot{x}_r + \beta (1 + x_r^2)\dot{x}_r + x_r - \frac{\Delta}{2}x_r = \alpha (\dot{x}_r + \dot{x}_l) \label{eq:mdl_vocal_fold_right} \\
    &\ddot{x}_l + \beta (1 + x_l^2)\dot{x}_l + x_l + \frac{\Delta}{2}x_l = \alpha (\dot{x}_r + \dot{x}_l) \label{eq:mdl_vocal_fold_left}
\end{align}
where $\alpha$ is the coupling coefficient between the supra- and sub-glottal pressure, $\beta$ incorporates mass, spring and damping coefficients of the vocal folds,  and $\Delta$ is an asymmetry coefficient. For a male adult with normal voice, their values (calculated from actual videographic measurements), average to around $\alpha \approx 0.25$, $\beta \approx 0.32$ and $\Delta \approx 0$.

The solution of the dynamical system above yields the displacement, velocity and acceleration of the vocal folds as a set of time-series.
The time-series corresponding to $x_r$ and $x_l$ represent the oscillations of the vocal folds. To obtain these, the \textit{forward problem} of estimating the time series must be jointly solved with the {\em inverse} problem of estimating the parameters of the dynamical system themselves. In \cite{wenbo}, we introduced the ADLES algorithm that achieves this by minimizing the error between the glottal flow waveform obtained by inverse filtering, and the vocal fold oscillations predicted by the model as its parameter space is sampled. This joint estimation algorithm is briefly explained in the section below.

\subsection{Solving the forward and inverse problems jointly}

During phonation, the vocal tract (of length $L$) acts as a filter that modulates the pressure wave produced by the airflow through the glottis: $\mathcal{F}: p_0(t) \mapsto p_L(t)$. $p_0(t)$, the pressure at the glottis,  can be deduced from $p_L(t)$, the pressure sensed by a microphone close to the lips, through inverse filtering: $p_0(t) = \mathcal{F}^{-1}(p_L(t))$. If $A(0)$ represents the cross-sectional area of the vocal channel at the glottis, then the volume velocity of airflow at the glottis, $u_0(t)$, can  be deduced from $p_0(t)$ at the glottis as $u_0^m(t) = \frac{A(0)}{\rho c}p_0(t)$, where $c$ is the speed of sound and $\rho$ is the ambient air density. The superscipt $m$ denotes that $u_0^m(t)$ is estimated from the pressure wave measured by a microphone near the mouth.

The volume velocity $u_0(t)$ can \textit{also} be estimated from the solution to the model in Eqns. \ref{eq:mdl_vocal_fold_left} and \ref{eq:mdl_vocal_fold_right}: $u_0(t) = \Tilde{c} d(2x_0 + x_l(t) + x_r(t))$, where $d$ is the length of vocal folds, and $\Tilde{c}$ is the air particle velocity at the midpoint of the vocal fold. 

We derive our model parameters such that the glottal flow $u_0(t)$ predicted by the model matches the measured flow $u_0^m(t)$ as closely as possible. We define the {\em residual} $R(t) = u_0(t) - u_0^m(t)$ as the difference between the predicted and actual glottal flows, and the residual energy as 
\begin{equation}
    \mathcal{E} = \int_{0}^T R(t)^2 dt \label{eq:main_obj}
\end{equation}
We  estimate our model parameters to minimize the residual energy $\mathcal{E}$ subject to
Eqns. \ref{eq:mdl_vocal_fold_right} and \ref{eq:mdl_vocal_fold_left}, and  boundary constraints:
\begin{equation}
    x_r(0) = C_r, \, x_l(0) = C_l,\, \dot{x}_r(0) = 0, \, \dot{x}_l(0) = 0 \label{eq:boundary_conditions}
\end{equation}
where $C_r$ and $C_l$ are constants.  To solve the above functional least squares, we define the Lagrangian:
\begin{align}
\mathcal{L}  &= \mathcal{E} + \int_0^T (\lambda_r E_r + \lambda_l E_l) dt + \nu_l \dot{x}_l(0) + \nu_r \dot{x_r}(0) \nonumber \\
    &\quad  + \mu_l (x_l(0) - C_l) + \mu_r (x_r(0) - C_r)
    \label{eq:lagrangian}
\end{align}
where $E_r$ encodes the constraint of Eq. \ref{eq:mdl_vocal_fold_right}:
\begin{equation}
E_r = \ddot{x}_r + \beta (1 + x_r^2)\dot{x}_r + x_r - \frac{\Delta}{2}x_r - \alpha (\dot{x}_r + \dot{x}_l))
\end{equation}
and $E_l$ is similarly obtained from Eq. \ref{eq:mdl_vocal_fold_left}. $\lambda_l, \lambda_r, \mu_r, \mu_l, \nu_r$ and $\nu_l$ are Lagrangian multipliers. 
Differentiating $\mathcal{L}$ w.r.t. the model parameters and simplifying, 
% \footnote{https://arxiv.org/pdf/1910.08886.pdf}
%for $0 < t < T$ 
we get, for $\lambda_r$:
\begin{align}
    &\quad \ddot{\lambda}_r + (2\beta x_r\dot{x}_r + 1 - \frac{\Delta}{2})\lambda_r + 2\Tilde{c}dR = 0 \nonumber \\% \label{eq:adj_L}\\
    &\quad \beta(1 + x_r^2)\lambda_r - \alpha (\lambda_r + \lambda_l) = 0 \end{align}
and a similar pair of equations for $\lambda_l$ as well. At the end of the recording we also have: 
$$\lambda_r(T) = 0, \dot{\lambda_r}(T) = 0, \lambda_l(T) = 0, \dot{\lambda_l}(T) = 0 \label{eq:2_init_condition}$$ 
Substituting into the Lagrangian and simplifying we get the derivatives of $\mathcal{L}$ w.r.t. the model parameters:
\begin{align}
    &\quad \mathcal{L}_{\alpha} = \int_0^T - (\dot{x}_r + \dot{x}_l)(\lambda_r + \lambda_l) dt \label{eq:F_a} \\
    &\quad \mathcal{L}_{\beta} = \int_0^T ( (1 + x_r^2)\dot{x}_r\lambda_r + (1 + x_l^2)\dot{x}_l\lambda_l ) dt \label{eq:F_b}\\
    &\quad \mathcal{L}_{\Delta} = \int_0^T \tfrac1{2}(x_l\lambda_l - x_r\lambda_r) dt \label{eq:F_d}
\end{align}
Using gradient descent to optimize objective (\ref{eq:main_obj}), we get the following update rules:
\begin{align}
    \alpha^{k+1} = \alpha^{k} - \delta \mathcal{L}_{\alpha} \nonumber \\
    \beta^{k+1} = \beta^{k} - \delta \mathcal{L}_{\beta} \nonumber \\
    \Delta^{k+1} = \Delta^{k} - \delta \mathcal{L}_{\Delta} \label{eq:update_rule}
\end{align}
where $\delta$ is the step-size and $k$ refers to $k^{th}$ iteration.

\section{Experiments and results}\label{sec:3}

The algorithm described above is used to solve for the model parameters $\alpha$, $\beta$ and $\Delta$. These parameters are then substituted in the model to iteratively obtain $x_r$ and $x_l$.  The time series corresponding to $x_r$ and $x_l$ comprise the vocal fold oscillations. 
The behavior of their trajectories is studied in the model's phase space. The behavior can also be located on a bifurcation diagram that maps the behavior types in the model's parameter space. However, we do not extend our study to bifurcation diagrams in this paper. 

\textbf{Data used: } For our study we used a data set collected under clinical supervision and curated by Merlin Inc., a private firm in Chile. The dataset included recordings from 512 individuals who were tested for COVID-19, and turned out either COVID-19 postive or negative. Of these, we chose the recordings from only those individuals who had been recorded within 7 days of being medically tested. Only 19 individuals satisfied this criterion. Of these, 10 were females and 9 were males. 5 females and 4 males had been diagnosed with COVID-19, and the rest had tested negative.
The speech signals were sampled at 8 kHz, and recorded over microphones on commodity devices. Each individual was asked to utter multiple sounds, including the vowels /a/, /i/ and /u/. 

\textbf{Experiments performed: }
We performed two studies. In one, we estimated the vocal fold oscillations of the subjects in our dataset, observed the differences in the patterns of phase space trajectories of the model. Only the recordings of extended vowels /a/, /i/ and /u/ were used for this purpose. Each recording was sectioned into segments of 50ms duration, with an overlap of 25ms, generating 3835 sets of oscillation time-series in all.  We used the value of the residual $R(t)$ in Eq. \ref{eq:main_obj} to gauge our model's sufficiency in modeling extreme asymmetry in vocal fold motion. The value of $R(t)$ inversely relates to the accuracy with which the model is likely to estimate the vocal fold oscillations. %Very high values may indicate lower accuracy.
%The higher the value of $R$, the lower is this accuracy. 

In the second study, we used the residuals and the coefficients $\alpha$, $\beta$ and $\Delta$ as features, and investigated the use of several classifiers to discriminate between COVID-19 positive and negative individuals. The classifiers tested in this binary classification task were Logistic regression (LR), Support vector machine with a nonlinear radial basis function kernel
(NL-SVM), Decision tree (DT), Random forest (RF) tree and AdaBoost (AB). 3-fold cross validation experiments were done using recordings of the vowels /a/, /i/ and /u/.

% \begin{figure*}[!htb]
% \begin{minipage}[t]{0.24\linewidth}
% \centering \includegraphics[width=1.75in]{figs/Female_UPID-2743ea3a:split:vowel-i_UPID-2743ea3a_20200730121037_split.npy, sample-4000-4400, label- 0.png}
% \centering{(a) Female, negative}
% \end{minipage}
% \begin{minipage}[t]{0.24\textwidth}
% \centering \includegraphics[width=1.75in]{figs/Female_UPID-20ebc348:split:vowel-i_UPID-20ebc348_20200727160242_split.npy, sample-4000-4400, label- 1.png}
% \centerline{(b) Female, positive}
% \end{minipage}
% \begin{minipage}[t]{0.24\linewidth}
% \centering \includegraphics[width=1.75in]{figs/Male_UPID-54b7d518:split:vowel-i_UPID-54b7d518_20200728120755_split.npy, sample-4000-4400, label- 0.png}
% \centering{(a) Male, negative}
% \end{minipage}
% \begin{minipage}[t]{0.24\textwidth}
% \centering \includegraphics[width=1.75in]{figs/Male_UPID-2cf12bce:split:vowel-i_UPID-2cf12bce_20200702180743_split.npy, sample-4000-4400, label- 1.png}
% \centerline{(b) Male, positive}
% \end{minipage}
% \caption{Phase space trajectories for COVID-19 positive and negative individuals for the vowel /i/}
% \label{fig:phasorplots}
% \end{figure*}

\begin{figure*}[!htb]
\begin{minipage}[t]{0.24\linewidth}
\centering \includegraphics[width=1.75in]{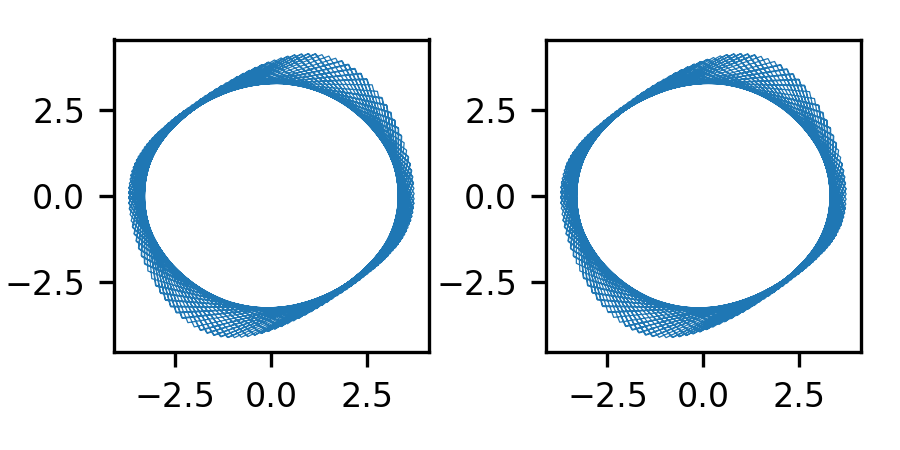}
\centering{(a) Female, negative}
\end{minipage}
\begin{minipage}[t]{0.24\textwidth}
\centering \includegraphics[width=1.75in]{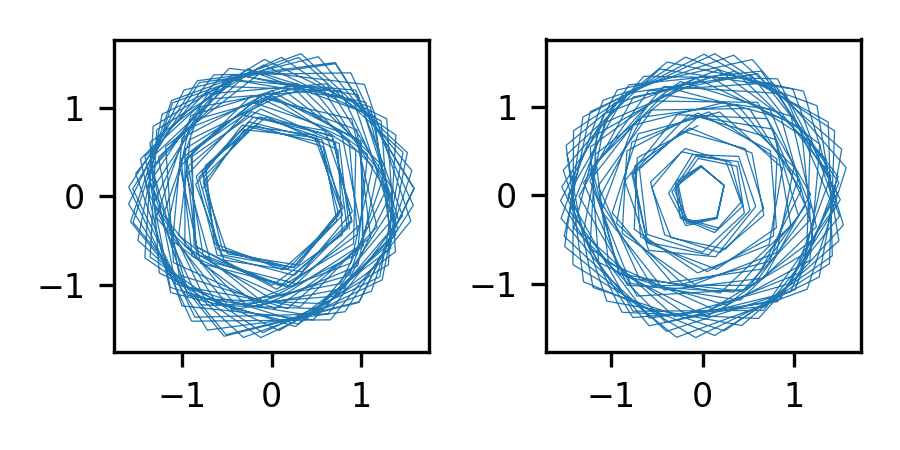}
\centerline{(b) Female, positive}
\end{minipage}
\begin{minipage}[t]{0.24\linewidth}
\centering \includegraphics[width=1.75in]{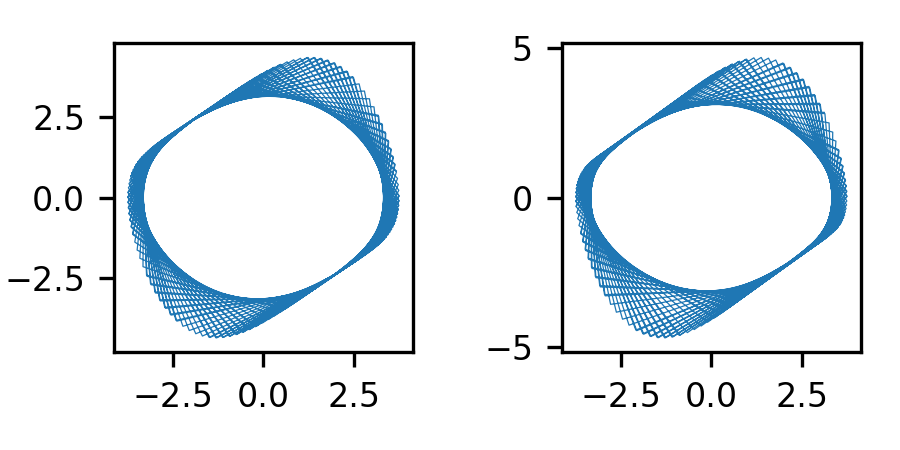}
\centering{(a) Male, negative}
\end{minipage}
\begin{minipage}[t]{0.24\textwidth}
\centering \includegraphics[width=1.75in]{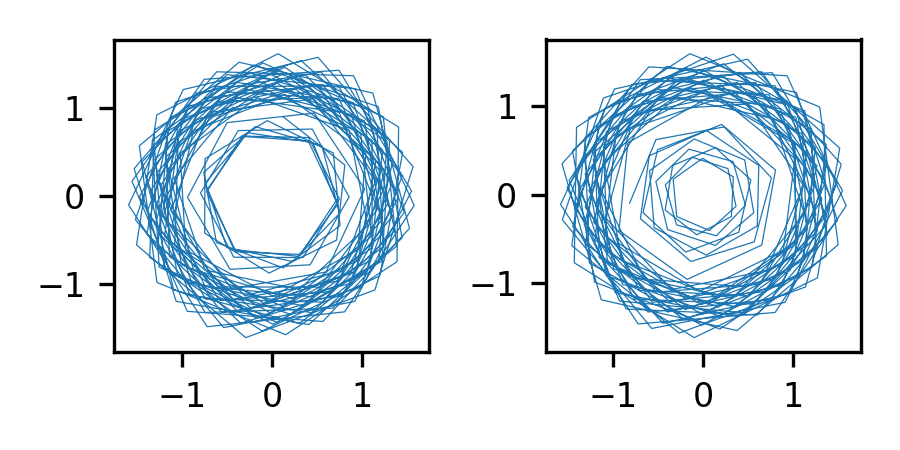}
\centerline{(b) Male, positive}
\end{minipage}
\caption{Phase space trajectories for the left and right vocal folds for COVID-19 positive and negative individuals for the vowel /i/. \textbf{Left panels:} $x_l$ ($x$-axis) vs. $\dot{x_l}$ ($y$-axis) \textbf{Right panels:} $x_r$ vs. $\dot{x_r}$ for each pair. }
\label{fig:phasorplots}
\vspace{-0.1in}
\end{figure*}

\begin{figure}[!htb]
\begin{minipage}[t]{0.4\linewidth}
\centering \includegraphics[width=3.4in, height=0.8in]{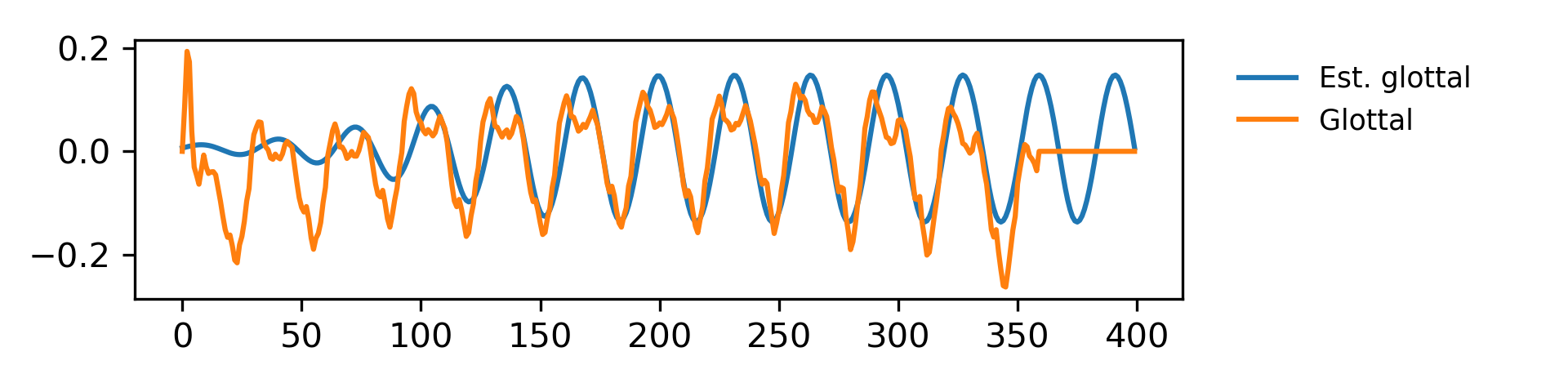}
\end{minipage}
\caption{Estimated vocal fold oscillations compared to the estimated glottal flow waveform of a subject}
\label{fig:oscillations}
\end{figure}

\textbf{Results of Study 1: }
The results of the first study are shown in Figs. \ref{fig:phasorplots} and \ref{fig:oscillations}. Fig. \ref{fig:phasorplots} shows the phase space trajectories of the model on a displacement vs. velocity plane for each vocal fold, for COVID-19 positive and negative patients of both genders. We see a significant difference in the phase space behaviors of COVID-19 positive and negative individuals (with a very small number of outliers the need to be investigated in further studies). The phase space trajectories for COVID-19 negative individuals are limit cycles or slim toroids, indicating a greater degree of synchronization in the eigenmodes of vibration, and greater symmetry of motion. For  COVID-19 positive patients, the trajectories are more complex, indicating a higher degree of both asynchrony and asymmetry and the \textsl{range of motion is reduced}. The vocal folds are unable to maintain the natural  self-sustained vibrations required for vocalization, and thier range of motion is restricted by an order of magnitude relative to normal. Although measures of divergence may be used to quantify these, e.g. Lyapunov exponents \cite{wolf1985determining}, we have not used these yet. 

Fig. \ref{fig:oscillations} shows a comparison of the estimated oscillations of the vocal folds to the glottal flow waveform obtained by inverse filtering. Note that in reality, the two are not the same. The former are the actual displacements of the vocal folds during phonation, the latter is the airflow volume velocity values across the glottis. Their strong correlation is however reflected in the example shown in Fig. \ref{fig:oscillations}. 

% \begin{figure}[!htb]
% \begin{minipage}[t]{0.4\linewidth}
% \centering \includegraphics[width=1.3in]{figs/_2800-3200__x_vs_x__8e2f1223_noncovid_female.png}
% \centering{(a) Female, negative}
% \end{minipage}
% \begin{minipage}[t]{0.4\textwidth}
% \centering \includegraphics[width=1.3in]{figs/_4000-4400__x_vs_x_20ebc348.png}
% \centerline{(b) Female, positive}
% \end{minipage}
% \\
% \begin{minipage}[t]{0.4\linewidth}
% \centering \includegraphics[width=1.3in]{figs/_2800-3200__x_vs_x__8e2f1223_noncovid_female.png}
% \centering{(a) Male, negative}
% \end{minipage}
% \begin{minipage}[t]{0.4\textwidth}
% \centering \includegraphics[width=1.3in]{figs/_4000-4400__x_vs_x_20ebc348.png}
% \centerline{(b) Male, positive}
% \end{minipage}
% \caption{Phase space trajectories for COVID-19 positive and negative individuals for the vowel /i/}
% \label{fig:phasorplots}
% \end{figure}

\textbf{Results of Study 2: } The results of the second study are shown in Tables \ref{table:all_vowels} and \ref{table:each_vowel}, In all experiments, performance was evaluated using the coresponding Reciever Operating Characteristics (ROC) curve. Tables \ref{table:all_vowels} and \ref{table:each_vowel} report the area under this curve (ROC-AUC) and its standard deviation (STD) for each experiment. % and Fig. \ref{fig:ROC}.  %Fig. \ref{fig:ROC} shows the receiver operating characteristic (ROC) curves for the vowel /i/ obtained with different classifiers in a 3-fold cross validation experiment. In all experiments,  performance were evaluated using the area under this curve (ROC-AUC) and its standard deviation (STD).

% \begin{figure}[htb]
%       \centering
%       \centerline{\includegraphics[width=5cm]{figs/all_vowels_roc_fold_3.png}}
%     \caption{ROC plot for the COVID-19 detection task using different simple classifiers. The vowels /a/, /i/, and /u/ were used.}
%     \label{fig:ROC}
% \end{figure}
Table \ref{table:all_vowels} presents the ROC-AUC and STD obtained for the vowels - /a/, /i/ and /u/. The segments used in the 3-fold cross-validation experiment were stratified -- the speakers in the training set were not included in the test set. 
We observe from Table \ref{table:all_vowels} that all the classifiers achieve a comparable performance of $\approx$ 0.8 ROC-AUC. The statistical significance was tested for all classifiers and all were found to be significant, with $p$-values better than  $1e^{-5}$. This strongly indicates that the features (residual values and vocal fold oscillation coefficients) can indeed capture the anomalous vibrations of COVID-19 patients without using sophisticated modeling techniques such as neural networks. %Results from Fig \ref{fig:ROC} also support this.

\begin{table}[!htb]
        \centering
        \begin{tabular}{ | l | c | c | c | c | c | }
            \hline
            Classifiers & LR & NL-SVM & DT & RF & AB \\ 
            \hline
						ROC-AUC & 0.825 & 0.789 & 0.803 & 0.794 & 0.812\\
            STD & 0.032 & 0.037 &  0.081 & 0.060 & 0.064\\
            \hline
        \end{tabular}
        \caption{Performance of different classifiers in a stratified 3-fold cross-validation experiment. }
        \label{table:all_vowels}
				\vspace{-0.1in}
\end{table}

%
%\begin{table}[!htb]
        %\centering
        %\begin{tabular}{ | c | c | c | }
            %\hline
            %Classifiers         & ROC-AUC      & STD  \\ 
            %\hline
            %Logistic Regression & 0.825        & 0.032\\
            %Nonlinear SVM       & 0.789        & 0.037\\
            %Decision Tree       & 0.803        & 0.081\\
            %Random Forest       & 0.794        & 0.060\\
            %Adaboost            & 0.812        & 0.064\\
            %\hline
        %\end{tabular}
        %\caption{Performance of different classifiers in a stratified 3-fold cross-validation experiment.}
        %\label{table:all_vowels}
%\end{table}

In order to gain further insight into the importance of these features, we examined the  splits within the decision tree classifier specifically. We found that the residual $R$ is  consistently the most important feature, indicating that the vocal fold displacements themselves are highly discriminative for COVID-19. We point out here that while high residual values are discriminative, extreme values may occur because of the inability of the simple model used to model abnormally deviant oscillations. More sophisticated models must be used to to oversome this shortcoming, for better accuracy.

\begin{table}[!htb]
    \centering
    %\begin{tabular}{ | c | c | c |}
        %\hline
        %Vowel               & ROC-AUC        & STD  \\ 
        %\hline
        %/a/                 & 0.728          & 0.089\\
        %/i/                 & \textbf{0.912} & 0.023\\
        %/u/                 & 0.877          & 0.035\\
        %/a/ + /i/           & 0.728          & 0.089\\
        %/a/ + /u/           & 0.784          & 0.038\\
        %/u/ + /i/           & 0.901          & 0.023\\       
        %\hline
    %\end{tabular}
		
		    \begin{tabular}{ | c | c | c | c|c|c|c|}
        \hline
%Vowel
 & /a/ & /i/ & /u/ & /a/+/i/ & /a/+/u/ & /i/+/u/ \\
\hline
AUC & 0.653 & \textbf{0.912} & 0.877 & 0.728 & 0.784 & 0.901 \\
STD & 0.119 & 0.023 & 0.035 & 0.089 & 0.038 & 0.023 \\
\hline
    \end{tabular}

    \caption{Performance of logistic regression on extended vowels and their combinations.}
    \label{table:each_vowel}
\end{table}

Table \ref{table:each_vowel} shows the performance of logistic regression on different vowels and their combinations. We observe that the vowel /i/ (a high front vowel) consistently yields the best performance, followed by /u/ (a high back vowel) then /a/ (a low back vowel). This indicates that the ability to reach the higher frequency energy peaks during phonation is compromised due to COVID-19 infection. 

\vspace{-0.1in}
\section{Conclusions}
\label{sec:conclusion}
While vocal fold oscillation patterns can be indicative of COVID-19, two caveats \textbf{must} be noted: a) they are likely to be useful \textit{only} in symptomatic patients, and b) the exclusiveness of the anomalies observed to other respiratory conditions has \textbf{not} been tested. We can only say that COVID-19 disrupts the entrainment of the vocal folds during phonation, and causes asymmetries in their motion, and that these characteristics can yield discriminative features that can be used to detect COVID-19 with even simple classifiers. Furthermore, it seems possible to achieve a high ROC-AUC using just a single phonated sound (e.g. the vowel /i/). We hope that the techniques presented in this paper can help facilitate future work towards a simple and cheap alternative for the rapid detection of COVID-19, using more sophisticated models to better capture pathological vocal fold oscillations.

\vfill\pagebreak

\bibliographystyle{IEEEbib}
\bibliography{refs}

\end{document}